# High-pressure synthesis of Dirac materials: layered van der Waals bonded BeN$_4$ polymorph

Maxim Bykov,[*,1,2] Timofey Fedotenko,[3] Stella Chariton,[4] Dominique Laniel,[3] Konstantin Glazyrin,[5] Michael Hanfland,[6] Jesse S. Smith,[7] Vitali B. Prakapenka,[4] Mohammad F. Mahmood,[2] Alexander F. Goncharov,[1] Alena V. Ponomareva,[8] Ferenc Tasnádi,[9] Alexei I. Abrikosov,[10] Talha Bin Masood,[10] Ingrid Hotz,[10] Alexander N. Rudenko,[11,12,13] Mikhail I. Katsnelson,[12,13] Natalia Dubrovinskaia,[3,9] Leonid Dubrovinsky,[*,14] Igor A. Abrikosov[*,9]

[1] The Earth and Planets Laboratory, Carnegie Institution for Science, Washington, DC 20015, USA
[2] College of Arts and Science, Howard University, Washington, DC 20059, USA
[3] Material Physics and Technology at Extreme Conditions, Laboratory of Crystallography, University of Bayreuth, 95440 Bayreuth, Germany
[4] Center for Advanced Radiation Sources, University of Chicago, 60637 Chicago, Illinois, USA
[5] Photon Sciences, Deutsches Electronen Synchrotron (DESY), D-22607 Hamburg, Germany
[6] European Synchrotron Radiation Facility, 38043 Grenoble Cedex 9, France
[7] HPCAT, X-ray Science Division, Argonne National Laboratory, Argonne, IL 60439, USA
[8] Materials Modeling and Development Laboratory, National University of Science and Technology 'MISIS', 119049 Moscow, Russia
[9] Department of Physics, Chemistry and Biology (IFM), Linköping University, SE-58183 Linköping, Sweden
[10] Department of Science and Technology (ITN), Linköping University, Norrköping, Sweden
[11] Key Laboratory of Artificial Micro- and Nano-Structures of Ministry of Education and School of Physics and Technology, Wuhan University, Wuhan 430072, China
[12] Radboud University, Institute for Molecules and Materials, 6525AJ Nijmegen, The Netherlands
[13] Department of Theoretical Physics and Applied Mathematics, Ural Federal University, 620002 Ekaterinburg, Russia
[14] Bayerisches Geoinstitut, University of Bayreuth, 95440 Bayreuth, Germany

*Correspondence to: Maxim Bykov maks.byk@gmail.com, Leonid Dubrovinsky leonid.dubrovinsky@uni-bayreuth.de, Igor A. Abrikosov igor.abrikosov@liu.se

**Abstract**

High pressure chemistry is known to inspire the creation of unexpected new classes of compounds with exceptional properties. Here we employ laser-heated diamond anvil cell technique for synthesis of a Dirac material, BeN4. A triclinic phase of beryllium tetranitride *tr*-BeN4 was synthesized from elements at ~85 GPa. Upon decompression to ambient conditions, it transforms into a compound with atomic-thick BeN4 layers interconnected via weak van der Waals bonds consisting of polyacetylene-like nitrogen chains with conjugated π-systems and Be atoms in square-planar coordination. Theoretical calculations for a single BeN4 layer show that its electronic lattice is described by a slightly distorted honeycomb structure reminiscent of the graphene lattice and the presence of Dirac points in the electronic band structure at the Fermi level. The BeN4 layer, *i.e.* beryllonitrene, represents a qualitatively new class of 2D materials that can be built of a metal atom and polymeric nitrogen chains and host anisotropic Dirac fermions.

The use of non-conventional methods of materials synthesis or established synthesis techniques in a non-trivial way may lead to breakthrough discoveries that revolutionize many research fields. One of the most striking examples is graphene. It was considered as an interesting two-dimensional (2D) model, though predominantly theoretical, before high quality graphitic films were derived by Novoselov *et al.* by mechanical exfoliation of small mesas of highly oriented pyrolytic graphite using adhesive tape [1]. Graphene is a material with gapless Dirac cones near the Fermi level with a linear energy-momentum dispersion near the Dirac points, and exhibiting ultrahigh carrier mobility. While the Dirac cones in graphene and many other 2D compounds are isotropic, their anisotropy could result in anisotropic carrier mobility, making it possible to realize direction-dependent quantum devices and motivating intense search for materials systems hosting the anisotropic Dirac cones. Experimental evidences of anisotropic Dirac fermions have been reported in bulk stoichiometric PtTe2 single crystal [2] and in 2D material borophene [3]. However, the Dirac point in these materials is far from the Fermi energy. A cone with its apex located near the Fermi level has been observed in BaFe2As2 [4], though the anisotropy was weak. There were theoretical predictions for the organic conductor α-(BEDT-TTF)2I3 at high pressure [5] and a B2S honeycomb monolayer [6]. There are proposals to achieve desired anisotropy using mechanical stress, external periodic potentials, heterostructures, or by a strong laser field [5–11]. However, experimental realization of a material hosting strongly anisotropic Dirac fermions at the Fermi energy is limited to a few examples [12,13] and calls for novel synthetic approaches.

The simplest design route of novel 2D materials is based on the substitution of elements in a known 2D material with its neighboring elements in the periodic table, for example the replacement of the carbon atom in graphene by boron and nitrogen leads to *h*-BN. Even more structural diversity and possible anisotropy may be achieved by considering the second-order neighbors, *e.g.* the hypothetical compound $BeN_2$ would have the same number of valence electrons as graphene but is intrinsically anisotropic in nature. The search of new 2D materials commonly starts with selecting of a known 3D precursor phase, most often thermodynamically stable or very long-lived metastable at ambient conditions, typically with a layered crystal structure with hexagonal motives. Such precursors are not currently available for the Be-N system, although, several compounds in the Be-N system were explored theoretically as potential 2D materials [14–16].

Nitrides possess the largest thermodynamic scale of metastability in terms of the energy differences between stable and metastable structures (~190 meV/atom), which may allow the kinetic stabilization of potentially useful compounds [17]. The synthesis of nitrogen-rich compounds is complicated by the great stability of the dinitrogen molecule $N_2$: the nitrides tend to decompose at relatively low temperatures, that may be below the temperatures that are required for the synthesis. One way to overcome this problem is to use the high-pressure conditions [18–21]. Furthermore, high-pressure is a useful tool to affect the bonding types, coordination environments and physical properties of compounds, which may lead to completely unexpected results. For example, beryllium, known to be tetrahedrally coordinated in inorganic compounds, was recently reported to incrementally change its coordination number from four to six in the crystal structure of $CaBe_2P_2O_8$ under compression [22]. Thus, potentially, the high-pressure synthesis of a beryllium nitride with octahedrally coordinated Be may provide room for unexpected structural changes on decompression. In this Letter we report the exploration of this pathway through the investigation of beryllium polynitrides.

High-pressure high-temperature chemical reactions in the Be-N system were studied in laser-heated diamond anvil cells (LHDACs) using synchrotron powder and single-crystal X-ray diffraction (Methods, Table S1 [23]). The experiments led to a synthesis of several Be polynitrides (Fig. 1, S1, S2, Table S2 [23]). In particular, at 84 GPa and ~2000 K, laser-heating of Be and $N_2$ resulted in the synthesis of a compound with the chemical composition $BeN_4$ and a triclinic structure, further referred to as ***tr*-BeN₄**. The full crystallographic information for this compound is given in the Table S3 [23]. The Be atom is coordinated by six nitrogen atoms forming an elongated octahedron. The covalently bonded nitrogen atoms form infinite zigzag chains running along the [001] direction (Fig. 1a). If viewed along the [001] direction (Fig. 1e), the structure can be understood as consisting of slightly corrugated BeN$_4$ layers, interconnected with each other by long Be-N bonds (~1.83 Å). The $tr$-BeN$_4$ is isostructural with the FeN$_4$ compound, which was previously synthesized at ~106 GPa from Fe and N$_2$ [24,25]. The metal-nitrogen bonds in BeN$_4$ have a slightly more ionic character than in FeN$_4$, which leads to a more uniform electron distribution within the polymeric chains (Fig. S3 [23]).

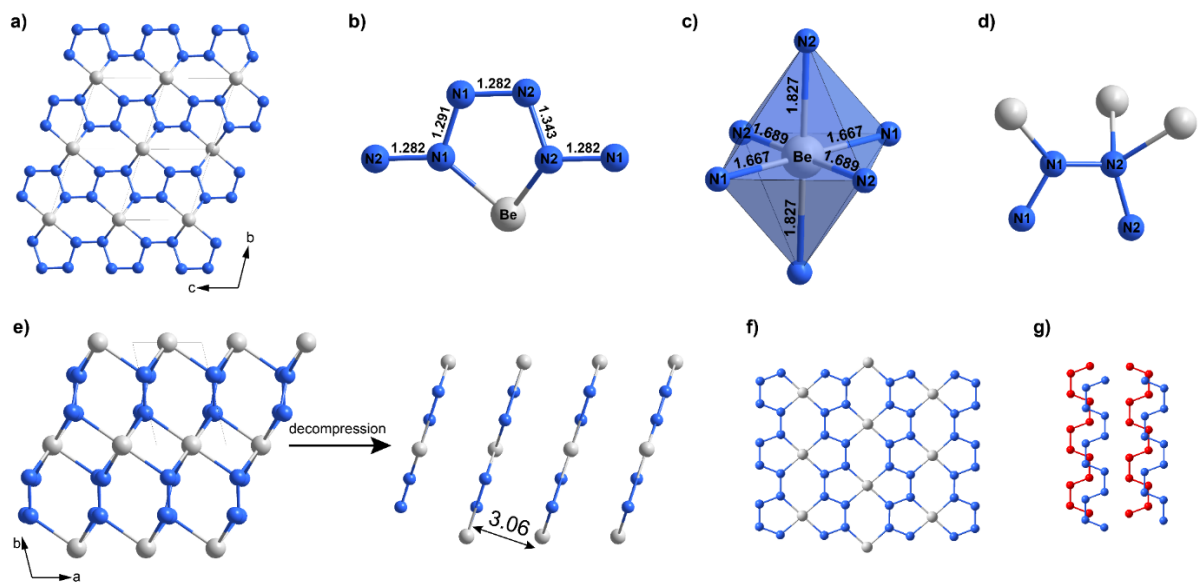

**Fig. 1. Crystal structure of $tr$-BeN$_4$ at 84 GPa.** (a) The view along the [100] direction; (b) coordination of a Be atom by one of the polymeric nitrogen chains. (c) BeN$_6$ octahedron with Be-N distances given in Å. (d) Coordination geometry of the nitrogen chain. (e) Scheme showing the transformation of $tr$-BeN$_4$ upon decompression. (f) Single layer of BeN$_4$ at ambient pressure. (g) Stacking sequence of polymeric nitrogen chains belonging to neighboring layers of $tr$-BeN$_4$ at ambient pressure; viewing direction is perpendicular to the layer. Beryllium and nitrogen atoms are represented as grey and blue spheres, respectively.

Theoretical calculations of the structural properties of $tr$-BeN$_4$ in a framework of density functional theory showed an excellent agreement between the theory and experiment for the lattice parameters, both at the synthesis pressure and on decompression (Fig. S4 [23]). Moreover, the structure is dynamically stable as indicated by the absence of the imaginary frequencies in the phonon dispersion calculations (Fig. S5 [23]). According to our calculated convex-hull diagram (Fig. S6 [23]) $tr$-BeN$_4$ remains thermodynamically stable down to at least 40 GPa and enters the metastability region upon further decompression.

On decompression the *tr*-BeN$_4$ produced excellent single-crystal diffraction patterns down to ~60 GPa, but at lower pressures the crystal quality started to deteriorate. Since the theoretical model perfectly matches the experimental structure at high pressures and the evolution of the strongest and well-separated (010) reflection well agrees with that suggested theoretically (Fig. S4d [23]), the evolution of the structure below 60 GPa was tracked by powder XRD and compared with the results of calculations. The weakening of the diffraction signal and the significant overlapping of the strongest reflections of *tr*-BeN$_4$ with those of ε-N$_2$ upon pressure decrease prevented us from unambiguous refinement of all six unit cell parameters at 20 and 34 GPa. However, the refinement was successfully made at ambient pressure (Figs. S4, S7 [23]) after the DAC was fully decompressed and nitrogen released. The decompressed sample of *tr*-BeN$_4$ is a highly textured powder and it contains α-Be$_3$N$_2$, which was not observed at higher pressures. Therefore, we conclude that at ambient conditions *tr*-BeN$_4$ slowly decomposes into α-Be$_3$N$_2$ but could potentially be completely kinetically stabilized at slightly lower temperature or by keeping the sample in a nitrogen atmosphere. No evidence of amorphous products (e.g. liquid nitrogen) in the diffraction pattern show that nitrogen was completely released from the DAC and the sample was already in contact with atmosphere.

The structural evolution of *tr*-BeN$_4$ on decompression involves breaking of two longer Be-N bonds in the BeN$_6$ octahedra which leads to a change of the coordination of the Be atom from octahedral to square planar. Consequently, all nitrogen atoms become $sp^2$-hybridised and form infinite planar polyacetylene-like chains (Fig. 2 and Movie S1 [23]). The process of rearrangement becomes apparent below 50 GPa, as it can be judged from the interplanar distance (Fig. S4 [23]). The evolution of the structure to the layered motif, which is dynamically stable at ambient pressure (Fig. S5b [23]), is accompanied by a significant depletion of the electron density between the layers (Fig. 2 and Movie S1 [23]). The topological data analysis (see Methods [23]) confirms that the charge density value corresponding to the layers' separation decreases with decreasing pressure. Thus, the BeN$_4$ layers are more weakly connected at ambient pressure than at high pressure, indicating the formation of a van der Waals-bonded solid, as confirmed by our theoretical analysis of the interlayer bonding in *tr*-BeN$_4$ and the exfoliation energy at ambient pressure (see Methods [23]).

Electronic structure calculations (Fig. S8 [23]) show that both the high pressure and the ambient pressure modifications of *tr*-BeN$_4$ are metallic. The calculated electron density maps and electron localization functions at a pressure of $P \sim 85$ GPa (Fig. S9 [23]) indicate a non-uniform distribution of electron density between the nitrogen atoms of the chains, in agreement with the crystal-chemical analysis (Fig. S3 [23]). During decompression, the electron density and interatomic distances become almost the same between all the atoms of the nitrogen chains, and the chains become planar (Fig. S9 [23]). Similar planar polyacetylene-like nitrogen chains with conjugated π-systems were recently observed in a series of metal-inorganic frameworks $Hf_4N_{22}$, $WN_{10}$, $ReN_{10}$, $Os_5N_{34}$ [26,27] and $MgN_4$ [28].

The high-pressure synthesis is a well-established path for the discovery of novel materials. A normal high-pressure synthesis route includes pressurizing and heating in the stability field of a new compound followed by its recovery to ambient pressure, hoping to preserve the material's interesting properties [29,30]. In this respect tr-BeN$_4$ is an exciting exception, as its layered ambient-pressure modification obtained upon decompression appears to possess even more intriguing properties than one could expect from the high pressure one. Its bulk electronic structure is modified in a remarkable fashion upon the decompression: one can clearly see the development of a pronounced pseudogap at the Fermi energy in tr-BeN$_4$ at low pressure. A closer examination of the bulk band structure and the Fermi surface (Fig. S8, Movie S2 [23]) allows us to conclude that there are two nodal points in the Brillouin zone (BZ) of tr-BeN$_4$ located at the Fermi energy. Given that both time-reversal and inversion symmetry are preserved, the two points are four-fold degenerate, indicating that they correspond to the Dirac points. The breaking of the inversion symmetry would split each of these points into Weyl points [31]. Though the electronic DOS in bulk tr-BeN$_4$ is finite, it originates from the well-defined Fermi surface pockets (electron and hole). Going from 3D to 2D, those pockets vanish, leaving only the Dirac points in the whole BZ. This makes the layered modification of tr-BeN$_4$ obtained upon decompression a precursor for a new anisotropic 2D Dirac material, the beryllonitrene. Our estimation of the exfoliation energy of the single monolayer for tr-BeN$_4$ (see Methods [23]) indicates that it is indeed possible.

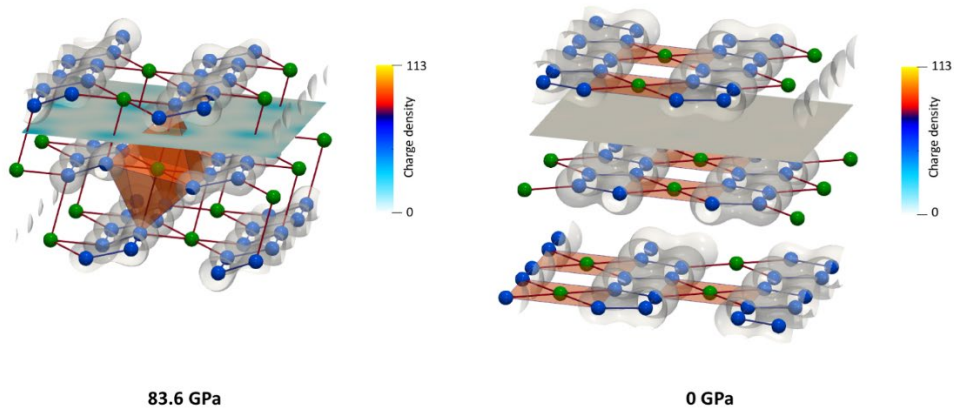

**Fig. 2. Evolution of the crystal structure and the charge density of *tr*-BeN₄ upon decompression from the synthesis pressure to ambient.** At the synthesis pressure, Be and N atoms form corrugated nets and the Be atoms (green) are octahedrally coordinated by six N atoms (blue) (left figure). Upon decompression, below about 50 GPa, the structure relaxes in such a way that the corrugated nets of Be and N atoms flatten, their increasing mutual separation reduces the coordination number of Be atoms to 4, thus turning the structure motif to layered (right figure). The grey planes between the layers indicate the position of the charge density slice and visualize the depletion of the charge density between two layers of *tr*-BeN₄ at zero pressure relative to that at synthesis pressure. Topological data analysis of the charge density (see Methods, Figs. S10-S11 [23]) confirms a stronger interaction within layers and weaker interaction between layers as the pressure is decreased. Experimental lattice parameters for *tr*-BeN₄ at 1 bar: $a$ = 3.275(2), $b$ = 4.212(2), $c$ = 3.704(2) Å, α = 103.43(3), β = 105.61(4), γ = 111.86(4)°, $V$ = 42.4 Å³. Optimized atomic positions: Be (0.5 0 0), N1 (0.1592 0.66170 0.5144), N2 (0.15840 0.6630 0.14780).

The main interest in fundamental studies and potential applications of 2D materials is related to their peculiar electronic properties. Therefore, we now turn to the electronic structure of the beryllonitrene. The optimized crystal structure of single-layer BeN₄ is shown in Fig. 3(a). It belongs to the oblique crystal system (the two-dimensional space group *p*2). The corresponding electronic band structure and density of states (DOS) are shown in Fig. 3(b). One can see that unlike bulk *tr*-BeN₄, the beryllonitrene is a Dirac semimetal with linear dispersion in the vicinity of the Fermi energy and two Dirac points coinciding with the Fermi energy, similar to graphene. Around the Fermi energy, electron and hole states are essentially symmetric over an energy range of roughly 2 eV. It is this property rather than the conical point itself which opens a way to deep relations to high energy physics within its symmetry between particles and antiparticles [32]. Unlike graphene, the Dirac points are not located at the corners of the BZ, but rather along the Γ-A direction (Σ point) as shown in the inset of Fig. 3(b). This can be attributed to a lower symmetry of the BeN₄ lattice compared to graphene. Indeed, one can see that the dispersion is different along the Σ-Γ and Σ-A directions, demonstrating a pronounced anisotropy of the band structure even in the long-wavelength limit.

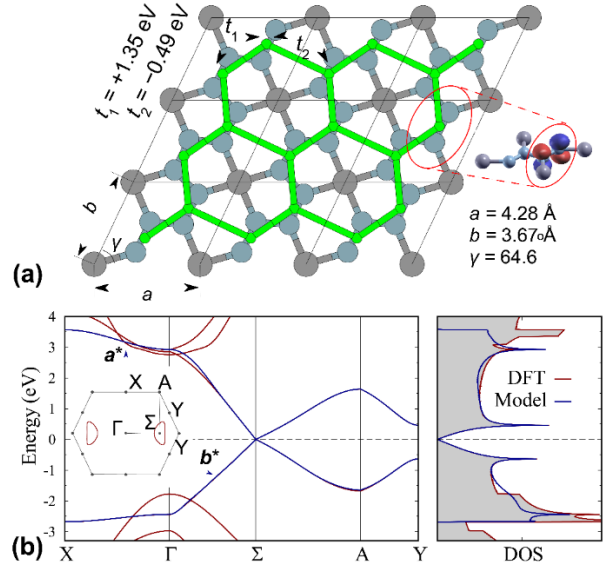

**Fig. 3**. **Electronic structure of the beryllonitrene** (a) Effective electronic lattice (green) superimposed onto the real-space crystal structure of single-layer BeN$_4$ (*a*). Green balls centered on the N-N bond correspond to the wave functions shown on the right side. Fourfold-coordinated gray balls depict Be atoms, dirty-blue three-fold coordinated balls correspond to N atoms. Arrows denote two main interaction parameters defined in terms of a tight-binding Hamiltonian. (b) Electronic band structure and the density of states calculated using DFT (red) and the *electronic* lattice model (blue) shown in (a). The inset shows the Brillouin zone with the high-symmetry points, as well as the Fermi surface corresponding to the Fermi energy of -0.5 eV

From the calculated band structure one can immediately estimate the carrier velocities in single-layer BeN$_4$ as $v(\mathbf{k}) = \frac{1}{\hbar}\frac{\partial E(\mathbf{k})}{\partial \mathbf{k}}$. Along the Σ-Γ direction, we find the Fermi velocity to be $v_y = 0.8\times10^6$ m/s, which is 1.3 times smaller than the Fermi velocity in graphene [33,34]. Along the Σ-A direction, the Fermi velocity is essentially smaller ($v_x = 0.3\times10^6$ m/s). At small charge doping the Fermi surface is elliptical with the dispersion given by $E(\mathbf{k}) = \hbar\left(v_x^2 k_x^2 + v_y^2 k_y^2\right)^{1/2}$. The anisotropy of the Fermi velocities is also reflected in the Fermi surface shown in the inset of Fig. 3(b). At sufficiently high doping, the Fermi surface is not elliptical, but is represented by two oppositely oriented semi-circles. It is worth noting that in this case the Fermi surface is perfectly nested by the wave vectors connecting two parallel parts of the two Fermi surface pockets. Therefore, the intervalley scattering in doped single-layer BeN$_4$ is expected to be strongly anisotropic.

To gain more insights into the physics of beryllonitrene, we construct a minimal lattice electronic model which describes its band structure (see Methods [23]). The corresponding "effective lattice" (electronic structure model) is shown in Fig. 3(a) superimposed onto the real-space crystal structure. It is described by a slightly distorted honeycomb net reminiscent to that of graphene. Each node of the honeycomb net is located at the center of the N-N bond. The model allows us to assess the relevance of screening effects in single-layer $BeN_4$. Calculations show (see Methods [23]) that at small wave vectors static dielectric function $\varepsilon \approx 8$ is notably larger compared to graphene (3.5), suggesting that the screening effects are more pronounced in berillonitrene. Another essential difference with graphene is the existence of van Hove singularities close enough to the Fermi energy (the separation is about 0.5 eV, contrary to approximately 2 eV in graphene). One can assume that this van Hove singularity is available by a reasonable doping; the Fermi energy shift by 0.5 eV is relatively easily reachable in graphene by chemical doping [35]. Recently, even much higher shift, about 2 eV, was reached in graphene by Gd intercalation [36]. In 2D systems with a Fermi energy close enough to the van Hove singularity one can expect the formation of flat bands and strong Fermi surface nesting. Due to many-body effects, this may result in all kind of correlation-induced instabilities (see Ref. [37] and refs. therein), including, e.g., the emergence of superconducting or magnetic phases [38] as well as charge- or spin-density waves [39].

To conclude, applying high-pressure synthesis followed by decompression to ambient conditions, we have synthesized new layered material, $BeN_4$. Van der Waals bonds between its layers and presence of anisotropic Dirac cones in its electronic structure show that 2D $BeN_4$, "beryllonitrene", has unique properties. Indeed, the high degree of electron-hole symmetry makes the 2D $BeN_4$ system similar, in some respect, to the world of high-energy particles with its symmetry between particles and antiparticles. However, contrary to the "true" Universe, which is isotropic, the massless Dirac fermions in the beryllonitrene are essentially anisotropic (see Methods [23]). This opens a door into the whole new world and the question arises on modification of "conventional" quantum relativistic effects such as Zitterbewegung, chiral tunneling, relativistic collapse at supercritical charges *etc*. [32] for the *anisotropic* massless fermions. We demonstrate the experimental realization of *tr*-$BeN_4$ in a diamond anvil cell and believe that further experiments aiming to control the quantity of the material or the crystallite size will allow the experimental realization of 2D beryllonitrene. Due to the relatively low nitrogen content in $BeN_4$ it potentially could be synthesized from beryllium azide $Be(N_3)_2$, which would not require a direct reaction between beryllium and nitrogen.

# References

[1] K. S. Novoselov, A. K. Geim, S. V. Morozov, Y. Jiang, S. V. Dubonos, I. V. Grigorieva, and A. A. Firsov, Science (80-. ). **306**, 666 (2004).

[2] M. Yan, H. Huang, K. Zhang, E. Wang, W. Yao, K. Deng, G. Wan, H. Zhang, M. Arita, H. Yang, Z. Sun, H. Yao, Y. Wu, S. Fan, W. Duan, and S. Zhou, Nat. Commun. **8**, 257 (2017).

[3] B. Feng, O. Sugino, R.-Y. Liu, J. Zhang, R. Yukawa, M. Kawamura, T. Iimori, H. Kim, Y. Hasegawa, H. Li, L. Chen, K. Wu, H. Kumigashira, F. Komori, T.-C. Chiang, S. Meng, and I. Matsuda, Phys. Rev. Lett. **118**, 096401 (2017).

[4] P. Richard, K. Nakayama, T. Sato, M. Neupane, Y.-M. Xu, J. H. Bowen, G. F. Chen, J. L. Luo, N. L. Wang, X. Dai, Z. Fang, H. Ding, and T. Takahashi, Phys. Rev. Lett. **104**, 137001 (2010).

[5] Y. Suzumura, J. Phys. Soc. Japan **85**, 053708 (2016).

[6] Y. Zhao, X. Li, J. Liu, C. Zhang, and Q. Wang, J. Phys. Chem. Lett. **9**, 1815 (2018).

[7] S.-M. Choi, S.-H. Jhi, and Y.-W. Son, Phys. Rev. B **81**, 081407 (2010).

[8] C.-H. Park, L. Yang, Y.-W. Son, M. L. Cohen, and S. G. Louie, Nat. Phys. **4**, 213 (2008).

[9] S. Katayama, A. Kobayashi, and Y. Suzumura, J. Phys. Soc. Japan **75**, 054705 (2006).

[10] M. Ezawa, Phys. Rev. Lett. **110**, 026603 (2013).

[11] C. Dutreix, E. A. Stepanov, and M. I. Katsnelson, Phys. Rev. B **93**, 241404 (2016).

[12] J. Kim, S. S. Baik, S. H. Ryu, Y. Sohn, S. Park, B.-G. Park, J. Denlinger, Y. Yi, H. J. Choi, and K. S. Kim, Science (80-. ). **349**, 723 (2015).

[13] X. Yuan, C. Zhang, Y. Liu, A. Narayan, C. Song, S. Shen, X. Sui, J. Xu, H. Yu, Z. An, J. Zhao, S. Sanvito, H. Yan, and F. Xiu, NPG Asia Mater. **8**, e325 (2016).

[14] C. Chen, B. Huang, and J. Wu, AIP Adv. **8**, 105105 (2018).

[15] C. Zhang and Q. Sun, J. Phys. Chem. Lett. **7**, 2664 (2016).

[16] Y. Ding, Y. Ji, H. Dong, N. Rujisamphan, and Y. Li, Nanotechnology **30**, 465202 (2019).

[17] W. Sun, S. T. Dacek, S. P. Ong, G. Hautier, A. Jain, W. D. Richards, A. C. Gamst, K. A. Persson, and G. Ceder, Sci. Adv. **2**, e1600225 (2016).

[18] B. A. Steele, E. Stavrou, J. C. Crowhurst, J. M. Zaug, V. B. Prakapenka, and I. I. Oleynik, Chem. Mater. **29**, 735 (2017).

[19] M. Bykov, K. R. Tasca, I. G. Batyrev, D. Smith, K. Glazyrin, S. Chariton, M. Mahmood, and A. F. Goncharov, Inorg. Chem. **59**, 14819 (2020).

[20] K. Niwa, T. Terabe, K. Suzuki, Y. Shirako, and M. Hasegawa, J. Appl. Phys. **119**, 065901 (2016).

[21] M. Bykov, K. V. Yusenko, E. Bykova, A. Pakhomova, W. Kraus, N. Dubrovinskaia, and L. Dubrovinsky, Eur. J. Inorg. Chem. **2019**, 3667 (2019).

[22] A. Pakhomova, G. Aprilis, M. Bykov, L. Gorelova, S. S. Krivovichev, M. P. Belov, I. A. Abrikosov, and L. Dubrovinsky, Nat. Commun. **10**, 2800 (2019).

[23] M. Supplemental, (n.d.).

[24] M. Bykov, E. Bykova, G. Aprilis, K. Glazyrin, E. Koemets, I. Chuvashova, I. Kupenko, C. McCammon, M. Mezouar, V. Prakapenka, H.-P. Liermann, F. Tasnádi, A. V. Ponomareva, I. A. Abrikosov, N. Dubrovinskaia, and L. Dubrovinsky, Nat. Commun. **9**, 2756 (2018).

[25] M. Bykov, S. Khandarkhaeva, T. Fedotenko, P. Sedmak, N. Dubrovinskaia, and L. Dubrovinsky, Acta Crystallogr. Sect. E Crystallogr. Commun. **74**, 1392 (2018).

[26] M. Bykov, S. Chariton, E. Bykova, S. Khandarkhaeva, T. Fedotenko, A. V. Ponomareva, J. Tidholm, F. Tasnádi, I. A. Abrikosov, P. Sedmak, V. Prakapenka, M. Hanfland, H.-P. Liermann, M. Mahmood, A. F. Goncharov, N. Dubrovinskaia, and L. Dubrovinsky, Angew. Chemie Int. Ed. **59**, 10321 (2020).

[27] M. Bykov, E. Bykova, E. Koemets, T. Fedotenko, G. Aprilis, K. Glazyrin, H.-P. P. Liermann, A.

V. Ponomareva, J. Tidholm, F. Tasnádi, I. A. Abrikosov, N. Dubrovinskaia, and L. Dubrovinsky, Angew. Chemie Int. Ed. **57**, 9048 (2018).
[28] D. Laniel, B. Winkler, E. Koemets, T. Fedotenko, M. Bykov, E. Bykova, L. Dubrovinsky, and N. Dubrovinskaia, Nat. Commun. **10**, 4515 (2019).
[29] M. Bykov, S. Chariton, H. Fei, T. Fedotenko, G. Aprilis, A. V. Ponomareva, F. Tasnádi, I. A. Abrikosov, B. Merle, P. Feldner, S. Vogel, W. Schnick, V. B. Prakapenka, E. Greenberg, M. Hanfland, A. Pakhomova, H.-P. Liermann, T. Katsura, N. Dubrovinskaia, and L. Dubrovinsky, Nat. Commun. **10**, 2994 (2019).
[30] P. F. McMillan, Nat. Mater. **1**, 19 (2002).
[31] S. M. Young, S. Zaheer, J. C. Y. Teo, C. L. Kane, E. J. Mele, and A. M. Rappe, Phys. Rev. Lett. **108**, 140405 (2012).
[32] M. I. Katsnelson, *The Physics of Graphene*, 2nd ed. (Cambridge University Press, Cambridge, 2020).
[33] K. S. Novoselov, A. K. Geim, S. V. Morozov, D. Jiang, M. I. Katsnelson, I. V. Grigorieva, S. V. Dubonos, and A. A. Firsov, Nature **438**, 197 (2005).
[34] Y. Zhang, Y.-W. Tan, H. L. Stormer, and P. Kim, Nature **438**, 201 (2005).
[35] R. R. Nair, I.-L. Tsai, M. Sepioni, O. Lehtinen, J. Keinonen, A. V. Krasheninnikov, A. H. Castro Neto, M. I. Katsnelson, A. K. Geim, and I. V. Grigorieva, Nat. Commun. **4**, 2010 (2013).
[36] S. Link, S. Forti, A. Stöhr, K. Küster, M. Rösner, D. Hirschmeier, C. Chen, J. Avila, M. C. Asensio, A. A. Zakharov, T. O. Wehling, A. I. Lichtenstein, M. I. Katsnelson, and U. Starke, Phys. Rev. B **100**, 121407 (2019).
[37] D. Yudin, D. Hirschmeier, H. Hafermann, O. Eriksson, A. I. Lichtenstein, and M. I. Katsnelson, Phys. Rev. Lett. **112**, 070403 (2014).
[38] V. Y. Irkhin, A. A. Katanin, and M. I. Katsnelson, Phys. Rev. B **64**, 165107 (2001).
[39] P. Monceau, Adv. Phys. **61**, 325 (2012).

## Acknowledgments

Parts of this research were carried out at the Extreme Conditions Beamline (P02.2) at DESY, a member of Helmholtz Association (HGF). Portions of this work were performed on beamline ID15 at the European Synchrotron Radiation Facility (ESRF), Grenoble, France. Portions of this work were performed at GeoSoilEnviroCARS (The University of Chicago, Sector 13) and at HPCAT (sector 16) of the Advanced Photon Source (APS), Argonne National Laboratory. Research was sponsored by the Army Research Office and was accomplished under the Cooperative Agreement Number W911NF-19-2-0172. N.D. and L.D. thank the Deutsche Forschungsgemeinschaft (DFG projects DU 954-11/1 and, DU 393-9/2, and DU 393-13/1) and the Federal Ministry of Education and Research, Germany (BMBF, grant no. No. 05K19WC1) for financial support. D.L. thanks the Alexander von Humboldt Foundation for financial support. Theoretical analysis of chemical bonding was supported by the Russian Science Foundation (Project No. 18-12-00492). Calculations of the phonon dispersion relations were supported by the Ministry of Science and Higher Education of the Russian Federation in the framework of Increase Competitiveness

Program of NUST MISIS (No. K2-2020-026) implemented by a governmental decree dated 16 March 2013, No. 211. Support from the Knut and Alice Wallenberg Foundation (Wallenberg Scholar Grant No. KAW-2018.0194), the Swedish Government Strategic Research Areas in Materials Science on Functional Materials at Linköping University (Faculty Grant SFO-Mat-LiU No. 2009 00971) and SeRC, the Swedish Research Council (VR) grant No. 2019-05600 and the VINN Excellence Center Functional Nanoscale Materials (FunMat-2) Grant 2016–05156 is gratefully acknowledged. The computations were enabled by resources provided by the Swedish National Infrastructure for Computing (SNIC) partially funded by the Swedish Research Council through grant agreement no. 2016-07213. The work of MIK was supported by the JTC-FLAGERA Project GRANSPORT. GeoSoilEnviroCARS is supported by the National Science Foundation – Earth Sciences (EAR – 1634415) and Department of Energy-Geosciences (DE-FG02-94ER14466). HPCAT operations are supported by DOE-NNSA's Office of Experimental Sciences. Advanced Photon Source is U.S. Department of Energy (DOE) Office of Science User Facility operated for the DOE Office of Science by Argonne National Laboratory under Contract No. DE-AC02-06CH11357.

# Supplementary Materials

# High-pressure synthesis of Dirac materials: layered van der Waals bonded BeN$_4$ polymorph

Maxim Bykov,[*,1,2] Timofey Fedotenko,[3] Stella Chariton,[4] Dominique Laniel,[3] Konstantin Glazyrin,[5] Michael Hanfland,[6] Jesse S. Smith,[7] Vitali B. Prakapenka,[4] Mohammad F. Mahmood,[2] Alexander F. Goncharov,[1] Alena V. Ponomareva,[8] Ferenc Tasnádi,[9] Alexei I. Abrikosov,[10] Talha Bin Masood,[10] Ingrid Hotz,[10] Alexander N. Rudenko,[11,12,13] Mikhail I. Katsnelson,[12,13] Natalia Dubrovinskaia,[3,9] Leonid Dubrovinsky,[*,14] Igor A. Abrikosov[*,9]

[1] The Earth and Planets Laboratory, Carnegie Institution for Science, Washington, DC 20015, USA
[2] College of Arts and Science, Howard University, Washington, DC 20059, USA
[3] Material Physics and Technology at Extreme Conditions, Laboratory of Crystallography, University of Bayreuth, 95440 Bayreuth, Germany
[4] Center for Advanced Radiation Sources, University of Chicago, 60637 Chicago, Illinois, USA
[5] Photon Sciences, Deutsches Electronen Synchrotron (DESY), D-22607 Hamburg, Germany
[6] European Synchrotron Radiation Facility, 38043 Grenoble Cedex 9, France
[7] HPCAT, X-ray Science Division, Argonne National Laboratory, Argonne, IL 60439, USA
[8] Materials Modeling and Development Laboratory, National University of Science and Technology 'MISIS', 119049 Moscow, Russia
[9] Department of Physics, Chemistry and Biology (IFM), Linköping University, SE-58183 Linköping, Sweden
[10] Department of Science and Technology (ITN), Linköping University, Norrköping, Sweden
[11] Key Laboratory of Artificial Micro- and Nano-Structures of Ministry of Education and School of Physics and Technology, Wuhan University, Wuhan 430072, China
[12] Radboud University, Institute for Molecules and Materials, 6525AJ Nijmegen, The Netherlands
[13] Department of Theoretical Physics and Applied Mathematics, Ural Federal University, 620002 Ekaterinburg, Russia
[14] Bayerisches Geoinstitut, University of Bayreuth, 95440 Bayreuth, Germany

*Correspondence to: Maxim Bykov maks.byk@gmail.com, Leonid Dubrovinsky leonid.dubrovinsky@uni-bayreuth.de, Igor A. Abrikosov igor.abrikosov@liu.se

## Methods

<u>Synthesis</u>

In every experiment the Be metal was placed in the sample chamber of a BX90 diamond anvil cell equipped with Boehler-Almax type diamonds (Re gasket, 120-250 μm culet diameter). The DACs were loaded with molecular nitrogen which served as a pressure-transmitting medium and as a reagent. The DACs were compressed up to the target pressures and heated with the double-sided laser-heating systems of the P02.2 (Petra III, DESY, Hamburg, Germany) and 13IDD (APS, Argonne, USA) beamlines. The summary of the heating experiments is shown in the Table S1.

<u>X-ray diffraction</u>

XRD measurements were performed at the beamlines P02.2 beamline of Petra III (DESY, Hamburg, Germany, $\lambda$ = 0.2891 Å, 2×2 μm$^2$ focusing, Perkin Elmer XRD1621 detector), the 13IDD beamline (APS, Argonne, USA, $\lambda$ = 0.2952 Å, 3×3 μm$^2$, CdTe Pilatus 1M detector), the 16-IDB beamline (APS, Argonne, USA, $\lambda$ = 0.34453 Å, 3×3 μm$^2$, Pilatus 1M detector) and the ID15B (ESRF, Grenoble, France, $\lambda$ = 0.4117 Å, 10×10 μm$^2$, MAR555 flat panel detector). For the single-crystal XRD measurements, the samples were rotated around a vertical ω-axis in the ±35° range. The diffraction images were collected with an angular step Δω = 0.5° and an exposure time of 5s or 10s/frame. For the analysis of the single-crystal diffraction data (indexing, data integration, frame scaling and absorption correction) we used the *CrysAlisPro* software package. To calibrate an instrumental model in the *CrysAlisPro* software, i.e., the sample-to-detector distance, detector's origin, offsets of goniometer angles, and rotation of both X-ray beam and the detector around the instrument axis, we used a single crystal of orthoenstatite ((Mg$_{1.93}$Fe$_{0.06}$)(Si$_{1.93}$, Al$_{0.06}$)O$_6$, *Pbca* space group, $a$ = 8.8117(2), $b$ = 5.18320(10), and $c$ = 18.2391(3) Å). Powder diffraction measurements were performed either without sample rotation (still images) or upon continuous rotation in the ±20°ω range. The images were integrated to powder patterns with the DIOPTAS software [1]. Le-Bail fits of the diffraction patterns were performed with the Jana2006 and Topas 4.2 software packages [2]. The structures were solved with the ShelXT structure solution program [3] using intrinsic phasing and refined with SHELXL using Olex2 [4,5]. The equations of state (EOS) of the synthesised materials were obtained by fitting pressure-volume dependence data using EoSFit7-GUI software [6]. CSD 2005719-2005720 contain the supplementary crystallographic data for this paper. These data can be obtained free of charge from FIZ Karlsruhe via www.ccdc.cam.ac.uk/structures.

<u>First principles calculations of *tr*-BeN$_4$</u>

The ab-initio calculations were performed using the projector-augmented-wave (PAW) method [7] as implemented in the VASP code [8–10]. The sampling for BZ integrations was performed using the Gamma scheme with 18×18×18 k-point grids. In the calculation of the Fermi surface denser k-points sampling, 31×31×31 was employed. The energy cutoff for the plane waves included in the expansion of wave functions was set to 800 eV. For calculation of exchange–correlation energy we used the Perdew-Burke-Ernzerhof GGA (PBE) [11] and optB88b-vdW [12] methods with nonlocal contributions, where the van der Waals interactions are included and the exchange functional is optimized for the correlation part [13]. The convergence criterion for the electronic subsystem was chosen to be equal to $10^{-4}$ eV for two subsequent iterations, and the ionic relaxation loop within the conjugated gradient method was stopped when forces became of the order of $10^{-3}$ eV/Å. Finite displacement method was used to calculate the phonon dispersions of $tr$-BeN$_4$ at 0 GPa with Phonopy [14]. Converged phonon dispersions were achieved using a (4 × 4 × 4) supercell with 320 atoms.

Electronic properties of $tr$-BeN$_4$

Analysis of the electronic density of states of $tr$-BeN$_4$ (Fig. S8) demonstrates that at high pressure the material is metallic and the main contribution to the DOS at the Fermi level comes from $p_x$ electrons of nitrogen chains, forming π-system. Upon decompression the DOS becomes more localized, the electronic bands become narrower and small energy gaps and pseudogaps between states appear, including a pseudogap in the energy range from -6 to -5 eV due to the break of multicenter Be-N2-Be bonds. We can also observe the pronounced pseudogap at the Fermi energy due to the relaxation of the π-system.

Topological data analysis of BeN$_4$ charge density

To elucidate the possibility of the synthesis of a single layer of BeN$_4$ we investigated the nature of interlayer bonding of the layered $tr$-BeN$_4$ considering the evolution of the charge density between the layers upon decompression. From Fig. 2 and Movie S1 one sees that the charge density between the layers decreases significantly with decreasing pressure, and at ambient pressure the charge density becomes almost zero. To quantify the pressure evolution of the charge density, we performed a topological data analysis [15,16] of the valence charge density of $tr$-BeN$_4$ at seven different pressure values, 83.6 GPa, 40.5 GPa, 25.6 GPa, 17.8 GPa, 11.8 GPa, 6.8 GPa and 0 GPa. A 2x2x2 supercell was created from the unit cell. Merge tree based topological data analysis was done for the supercells at volumes corresponding to the above specified pressures with periodic boundary conditions. Let f:D→R be the charge density field. For a given charge density iso-value ρ, the super level set is defined as the set of all points p∈D with

$f(p) \geq \rho$. A merge tree tracks the changes in the topology of the super level set as the iso-value is decreased from high to low values. Thus, it captures the birth, merge and death events of the super level set components. The analysis allows us to identify exact iso values at which qualitative changes occur in the topology of the charge density and to classify the changes. Since we are using the valence charge density for our analysis, Be atoms that donate their valence electrons to the N atoms, become effectively invisible in the valence charge density plots. Thus, the analysis focuses on the behavior of charge density in associated with the N atoms.

Let us consider an example of the merge tree constructed at zero pressure. It captured following significant events in the topology of the charge density iso surfaces when charge density value is decreased from high to low (see Fig. S10).

1. The super-level set components are born at the maxima in the charge density (single point, Fig. S10a). These maxima are due to the high number of electrons that N atoms attract. If one would continue to lower the iso value, discrete iso surfaces form around N atoms (Fig. S10 b).
2. These discrete iso surfaces merge at the saddle points in the charge density that correspond to critical values at N-N bonds (according to Bader's definition [17]) and eventually enclose N chains (Fig. S10c).
3. The charge density enclosing N chains within the same $BeN_4$ layer continue to grow, and eventually merge into one charge density layer (Fig. S10 d,e).
4. All the charge density layers merge into a single connected component (Fig. S10f), meaning that we can no longer distinguish between different $BeN_4$ layers by just looking at the charge density at the given iso value.

Fig. S11a shows the obtained critical charge density iso values where super-level set components are born, while Fig. S11b shows the obtained critical charge density iso values where topological change in the connectivity of super level sets (points 3 and 4 above) happens. These values are plotted as a function of pressure so that one can observe the difference in topology of charge density as the pressure changes. Figure S11a shows that at synthesis pressure the distribution of charge is unequal between the N atoms with two different maxima meaning that one pair of N atoms getting a bit more electrons than the other pair, this of course can be an effect of the system being in a six coordination state and two N atoms interact with neighboring Be atoms from different layers and attract electrons from them. As the pressure decreases to ambient pressure the maxima for the different N atoms converge suggesting a uniform interaction with the electron giving Be atom and consequently better separation between $BeN_4$ layers.

Fig. S11b shows that the layer separation charge density iso value to decreases as the pressure decreases to ambient pressure, that is, the BeN$_4$ layers are more strongly connected at synthesis pressure. In contrast, we observe an increase in the chain separation values with decreasing pressure, which suggests the strengthening of interactions between N chains within a layer at ambient pressure. Combined, we observe a stronger interaction within layers and weaker interaction between layers as the pressure is decreased. In fact, at 83.6 GPa, the chain separation and layer separation occur at similar values of the chare density, suggesting a somewhat uniform distribution of electrons at synthesis pressure and a more discreet distribution at ambient pressure.

Van der Waals nature of interlayer bonding of layered *tr*-BeN$_4$ and exfoliation energy at ambient pressure

To elucidate the possibility of the synthesis of a single layer of BeN$_4$ we investigated the nature of interlayer bonding of the layered *tr*-BeN$_4$. First, we studied the importance of Van der Waals (VdW) interactions in calculations of the structural properties of BeN$_4$. To do so we compare the results of theoretical calculations within VdW-optB88 functional that includes the interactions and semi-local PBE calculations. The results show excellent agreement between PBE, vdW and experimental results at high pressure (Fig. S4a). However, as one sees in Fig. S4a upon decompression the agreement of the results with vdW correction with experiment becomes increasingly better in comparison with the semi-local PBE calculations. This provides additional argument that at ambient pressure we deal with a van der Waals material.

Second, we calculated the exfoliation energy, the energy required to remove one atomic layer from the surface of the bulk material, as the difference in the ground-state energy between a slab of N atomic layers and a slab of (N -1) atomic layers plus an atomic layer separated from the slab:

$$E_{exf} = -[E_{N-layer\ slab} - E_{(N-1)-layer\ slab} - E_{single\ layer}],$$

where $E$ - is the energy of supercells per surface atom.

Two series of calculations with supercells N=10 and N=15 give $E_{exf}$= 79.8 meV/atom and 79.3 meV/atom, respectively. We find that the vdW results for the *tr*- BeN$_4$ exfoliation energy are in the range of values corresponding to the experimental [18,19] and theoretical [20] results for the graphite exfoliation energy.

First-principles calculations of single-layer BeN$_4$

Calculations were carried out using density functional theory (DFT) within the projected augmented wave method [7] as implemented in the *Vienna ab initio simulation package* (VASP) [9,10]. A generalized gradient approximation [11] was employed to describe exchange-correlation effects. The kinetic energy cutoff was set to 400 eV and the BZ sampled by a (14×16) $k$-point mesh. The vacuum thickness of 25 Å was used to avoid spurious interactions between the supercell periodic images in the direction perpendicular to the 2D plane. An energy window of at least 50 eV was used in the polarizability calculations. The construction of the Wannier functions and tight-binding parametrization of the DFT Hamiltonian was performed using the maximal localization procedure [21] by means of the Wannier90 code [22].

Minimal electronic model of BeN$_4$ band structure

In our model calculations, the electronic lattice is described by a slightly distorted honeycomb structure reminiscent of the graphene lattice. Each lattice site is located at the center of the N-N bond Fig. 3a, and is represented by a Wannier function in the form of a superposition of $p_z$-like orbitals, $|w\rangle \approx |p_z^1\rangle - |p_z^2\rangle$. The electronic properties of such effective lattice can be described by a tight-binding Hamiltonian $H = \sum_{ij} t_{ij} c_i^\dagger c_j$, where $t_{ij}$ is the hopping integral between lattice sites $i$ and $j$, and $c_i^\dagger (c_j)$ is the creation (annihilation) operator of an electron at site $i$ ($j$). The leading interaction terms correspond to the nearest- and next-nearest-neighbor terms with magnitudes $t_1$=1.35 eV and $t_2$=−0.49 eV, respectively. The difference in the hopping parameters is attributed to the lattice anisotropy. The constructed model accurately describes the electronic structure around the Fermi energy, as can be seen from Fig. 3(b).

To provide further details, we assess the relevance of screening effects in single-layer BeN$_4$. In the long-wavelength limit, the dielectric function of a pristine 2D Dirac system in Random Phase Approximation (RPA) reads $\varepsilon = 1 + \frac{\pi}{2}\alpha$, where $\alpha = \frac{e^2}{\hbar v}$ is an effective fine-structure constant [23]. Note that lattice Quantum Monte Carlo simulations [24] show that RPA results for the dielectric constant of massless Dirac fermions are pretty close to the exact ones. In anisotropic case we have the same expression but with $v = \sqrt{v_x v_y}$ [25]. Substituting our calculated velocities, we obtain $\varepsilon \approx 8$.

Beyond the long-wavelength limit, the screening effects can be quantified by the strength of the effective Coulomb interaction defined in terms of the Hubbard model [23]. Practically, it can be estimated within the constrained random phase approximation (cRPA) [26]. The cRPA assumes static dielectric screening of the bare Coulomb interaction $V(r_{ij})$ between Wannier orbitals defined in terms of the tight-binding model by all high-energy states non included in the model: $U(q)=[1-V(q)\Pi(q)]V(q)$, where $U(q)$ is the effective Coulomb interaction in $q$-space, and $\Pi(q)$ is the polarizability capturing relevant screening processes. Our estimate yields $U_0$=5.3 eV for the on-site interaction, and $U_1$=3.8 eV and $U_2$=2.2 eV for the nearest- and next-nearest-neighbor interactions, respectively. The obtained interactions are up to two times smaller compared to graphene [27]. At the same time, the half-bandwidth in single-layer BeN$_4$

($W$=3.1 eV) is comparable to that in graphene (2.8 eV). The degree of the local electronic correlations can be estimated by the ratio $U_0/W$. We arrive at the following upper limit $U_0/W \sim 1.7$, which suggests that single-layer BeN$_4$ is less correlated material compared to graphene. Particularly, pristine single-layer BeN$_4$ is expected to be further away from the Mott metal-insulator transition, magnetic instabilities, and other exotic quantum effects [28].

Massless Dirac particles with Coulomb interaction

For massless Dirac particles with Coulomb interaction the Fermi velocity is renormalized by many-body corrections logarithmically divergent at $k_F \to 0$, $k_F$ is the Fermi wave vector [23,29]. For the case of anisotropic massless Dirac fermions these corrections were calculated in Ref. [25]. The result is:

$$\delta v_x = \frac{e^2}{\pi \hbar} \frac{v_x}{v_y} \left[ K(m) - \frac{v_x^2}{v_y^2} R(m) \right] \xi,$$

$$\delta v_y = \frac{e^2}{\pi \hbar} [K(m) - P(m)] \xi,$$

where $\xi = \ln \frac{1}{k_F a}$, $a$ is the lattice constant, $m = 1 - \frac{v_x^2}{v_y^2}$ (we assume $v_x \leq v_y$),

$$K(m) = \int_0^{\pi/2} \frac{d\theta}{\sqrt{1 - m\sin^2\theta}}$$

is the full elliptic integral of the first kind,

$$P(m) = \int_0^{\pi/2} \frac{d\theta \sin^2\theta}{(1 - m\sin^2\theta)^{3/2}}, \quad R(m) = \int_0^{\pi/2} \frac{d\theta \cos^2\theta}{(1 - m\sin^2\theta)^{3/2}}.$$

With our parameters, we have $\frac{\delta v_x}{v_x} = 1.9\xi$, $\frac{\delta v_y}{v_y} = -3.0\xi$. The difference in sign reflects the fact that renormalization tends to reduce the anisotropy of carrier velocities, in agreement with Ref. [25]. Keeping in mind that $k_F a \gg 1$ and $\xi \gtrsim 1$, the obtained corrections turn out to be larger than the order of the calculated velocities. On the one hand, this result means that the first-order perturbative corrections are insufficient to correctly describe the velocity renormalization. On the other hand, this suggests an important role of the many-body effects in the physics of anisotropic Dirac fermions in single-layer BeN$_4$. Anyway, the conclusion is that the many-body effects due to the long-range Coulomb interaction are much stronger in BeN$_4$ than in graphene where in the same approximation $\frac{\delta v}{v} \approx 0.5\xi$ [23].

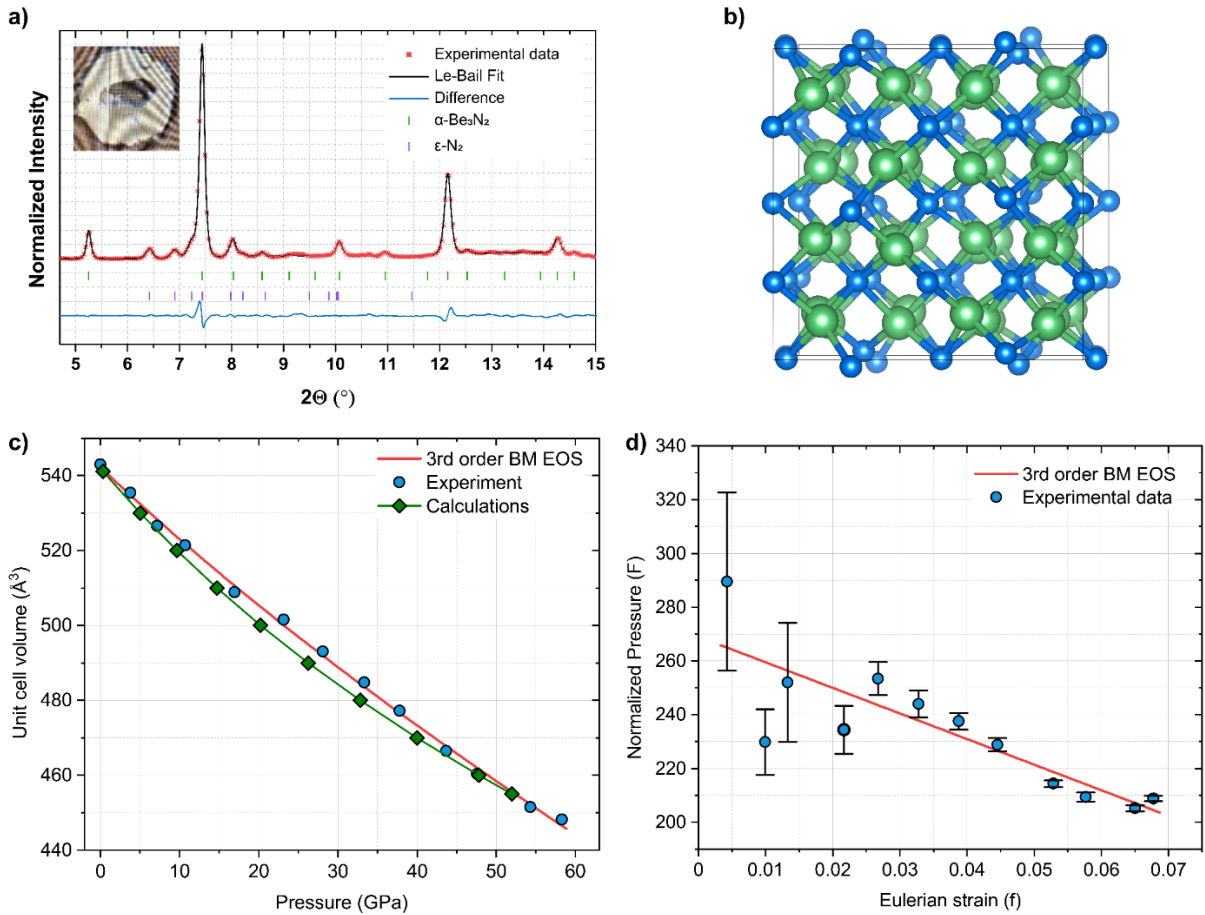

**Figure S1.** (a) Powder diffraction pattern of as-synthesised Be$_3$N$_2$ at 42 GPa ($\lambda = 0.29$ Å). (b) The crystal structure of α-Be$_3$N$_2$. Green and blue spheres represent the positions of Be and N atoms, respectively. The structure can be understood as a defect fluorite-type structure with distorted cubic closest packing of the nitrogen atoms, while the metal atoms occupy 75% of the tetrahedral holes in an ordered manner. The coordination pattern can be expressed as $^{3}_{\infty}[Be_3^{[4]}N_2^{[6]}]$, where the polyhedra around Be and around the nitrogen atoms are both distorted. (c) The pressure dependence of the unit cell volume of Be$_3$N$_2$. Experimental points are shown by circles. The solid curve is the fit of the experimental *P-V* data using the 3rd order Birch-Murnaghan EOS with the following parameters: $V_0$= 542.3($\pm$1.3) Å$^3$ $K_0$ = 269(14) GPa, and $K'_0$ = 1.6(3). Errors in experimental data are within the symbol size. Green symbols and points show the calculated EoS of Be$_3$N$_2$ obtained within vdW-optB88 functional. (d) F-f plot of α-Be$_3$N$_2$ based on the experimental data.

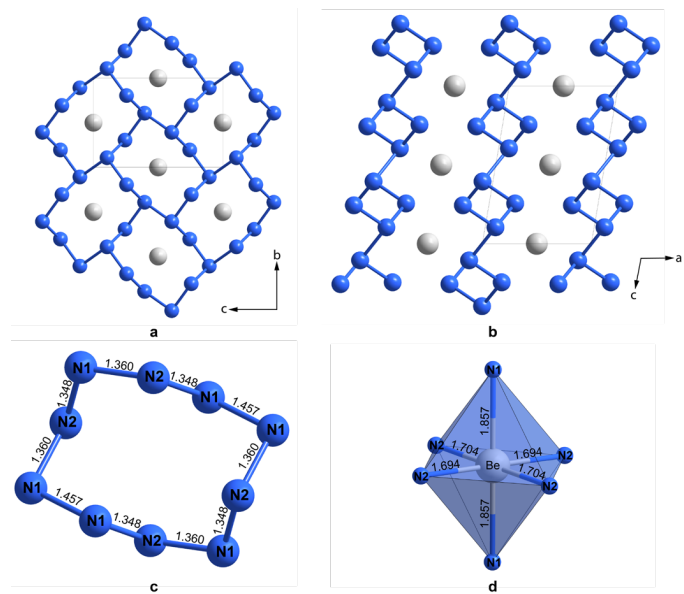

**Figure S2. Crystal Structure of *m*-BeN$_4$ at 98 GPa**. The interconnected nitrogen atoms form 2D networks constructed of condensed corrugated N$_{10}$ rings with all N atoms in tetrahedral coordination, while the Be atoms possess a distorted octahedral coordination (a) View along the [100] direction; (b) view along the [010] direction; (c) 10-member ring of N atoms with interatomic distances indicated in Å; (d) BeN$_6$ coordination polyhedron.

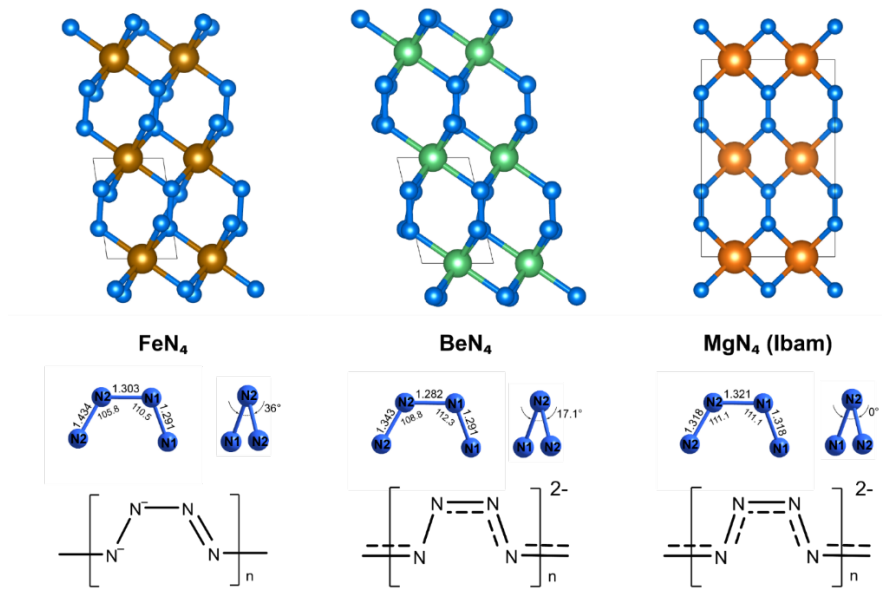

**Figure S3.** Comparison of the bonding within the polymeric nitrogen chains in a series of tetranitrides FeN$_4$, BeN$_4$ and MgN$_4$. On the first glance, the connectivity within the nitrogen chains of BeN$_4$ is similar to that in FeN$_4$: The N1 atom has a trigonal–planar coordination, while N2 is tetrahedrally coordinated, suggesting $sp^2$ and $sp^3$ hybridization, respectively. Therefore, the nitrogen atoms should form *catena*-poly[tetraz-1-ene-1,4-diyl] anions [–N–N–N=N–]$^{2-}_\infty$. However, the N chains in BeN$_4$ are much more planar than in FeN$_4$ with N2N2N1N1 torsion angle of 17.1° *vs* 36° in FeN$_4$. BeN$_4$ is in the intermediate situation between FeN$_4$ and MgN$_4$, [30,31] with the latter containing planar polyacetylene-like N chains with N2N2N1N1 angle of 0°. Polyacetylene-like chains with conjugated π-system and polytetrazene chains are isoelectronic with an ionic formula [N$_4$]$^{2-}$, however the polyacetylene chains are expected to be a more stable structural unit due to the stabilization by conjugation. Nevertheless, significant polar covalent character of the Fe-N bonds imposes geometric constraints, that leads to the redistribution of electron density in the nitrogen chains and to the formation of distorted chains with alternating double and single bonds. Indeed, the electronegativity of metal decreases in the row Fe – Be – Mg suggesting that the latter should form ionic compounds with nitrogen. Calculated Bader charges on Be and Fe in BeN$_4$ and FeN$_4$ (+1.77 and +1.22 respectively) support more ionic character of BeN$_4$. We therefore suggest that BeN$_4$ contains unique N-chains that represent the intermediate state between polytetrazene and polyacetylene-type chains.

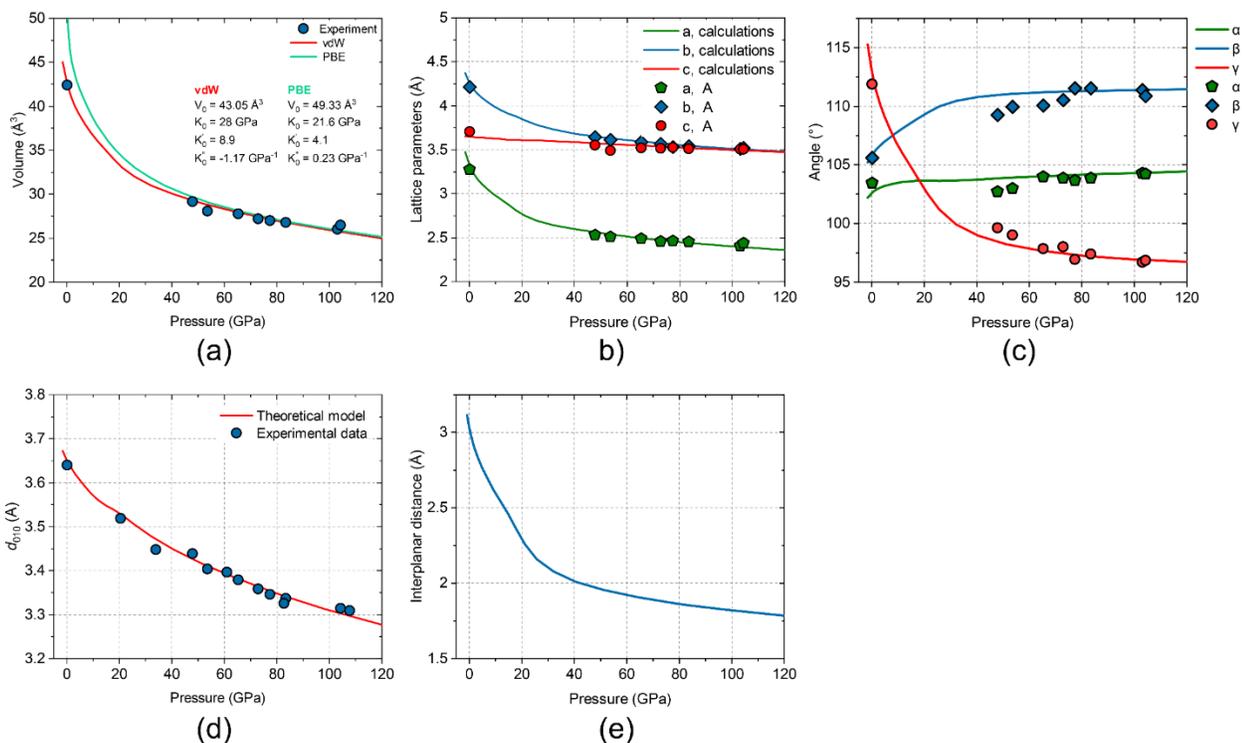

**Figure S4.** (a-c) Calculated and experimental lattice parameters of tr-BeN$_4$ as a function of pressure. (d) Pressure dependence of the d-spacing of the reflection (010) compared with the theoretical model. (e) Calculated d-d interplanar distance of *tr*-BeN$_4$ as a function of pressure.

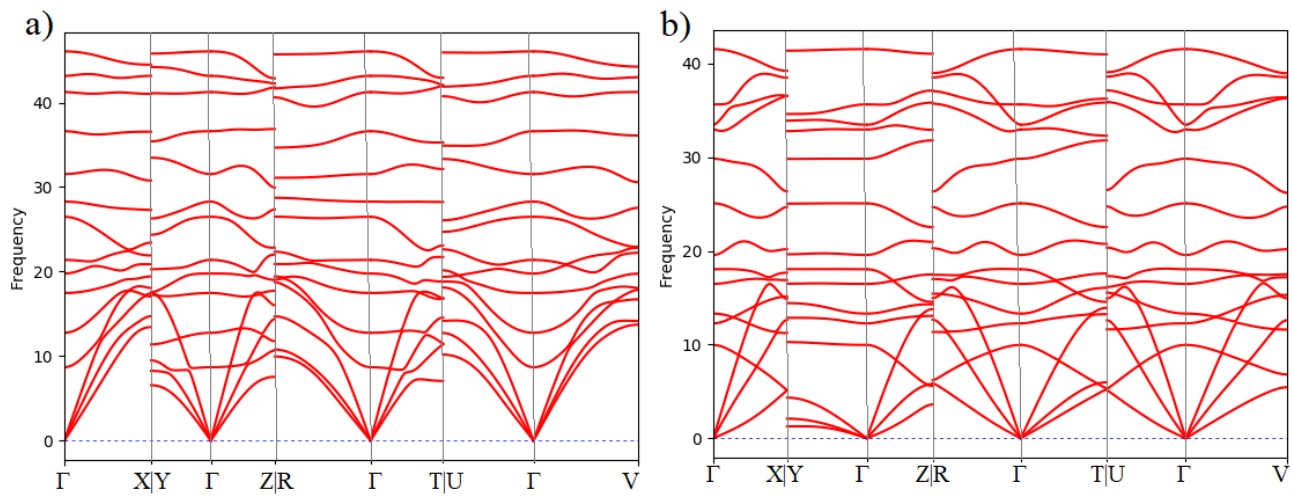

**Figure S5.** Phonon dispersion relations calculated for *tr*-BeN$_4$ at volumes 26.8 Å$^3$ (a) and 43.05 Å$^3$ (b), which correspond to pressures of P ~ 85 GPa and P ~ 0 GPa.

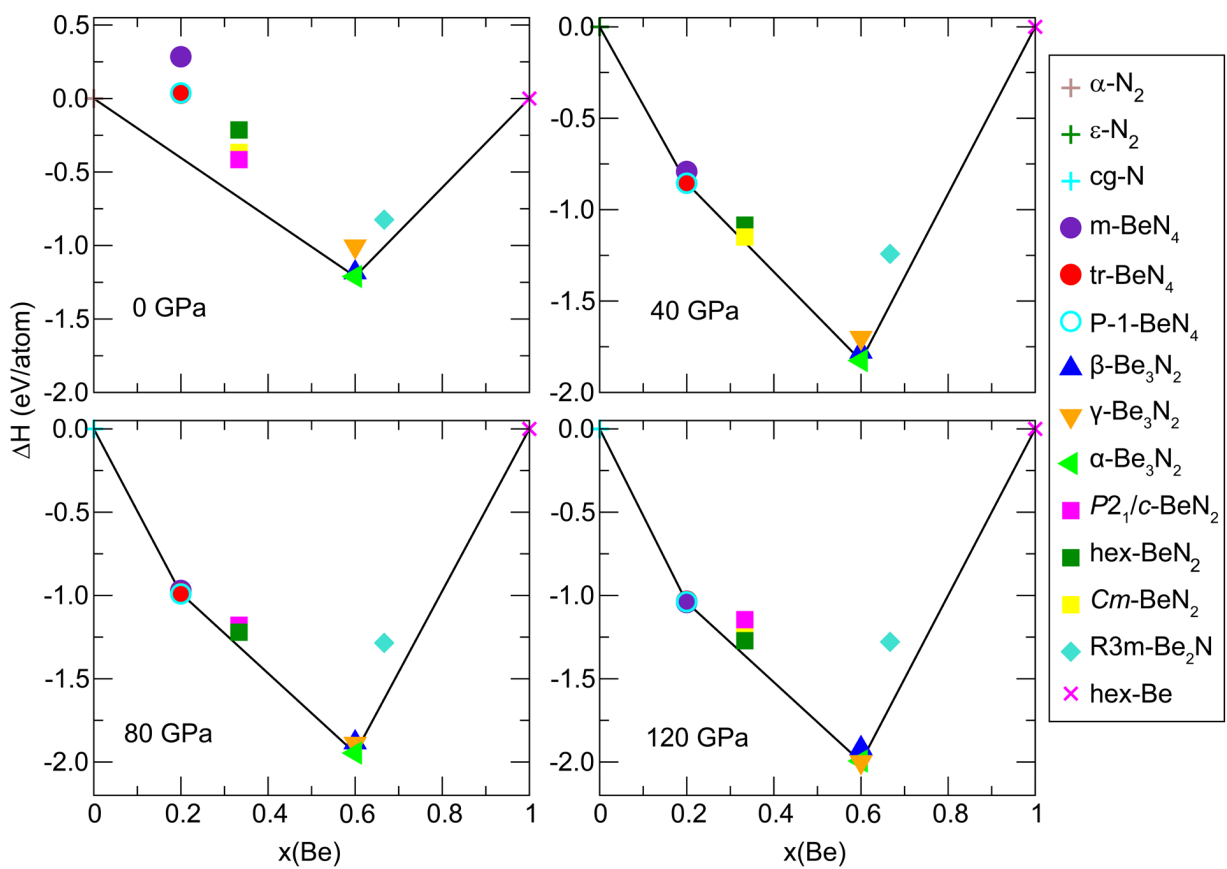

**Figure S6**. Convex hull diagram for Be-N system at various pressures. The considered phases include experimental phases from this study (*m*-BeN$_4$, *tr*-BeN$_4$, α-Be$_3$N$_2$) and previously predicted or synthesized phases (α-N$_2$ (0 GPa) [32], ε-N$_2$ (40 GPa) [33], cg-N (80 and 120 GPa) [34], P-1 BeN$_4$ [35], BeN$_2$ [36], β- and γ-Be$_3$N$_2$ [37] and Be$_2$N [38]).

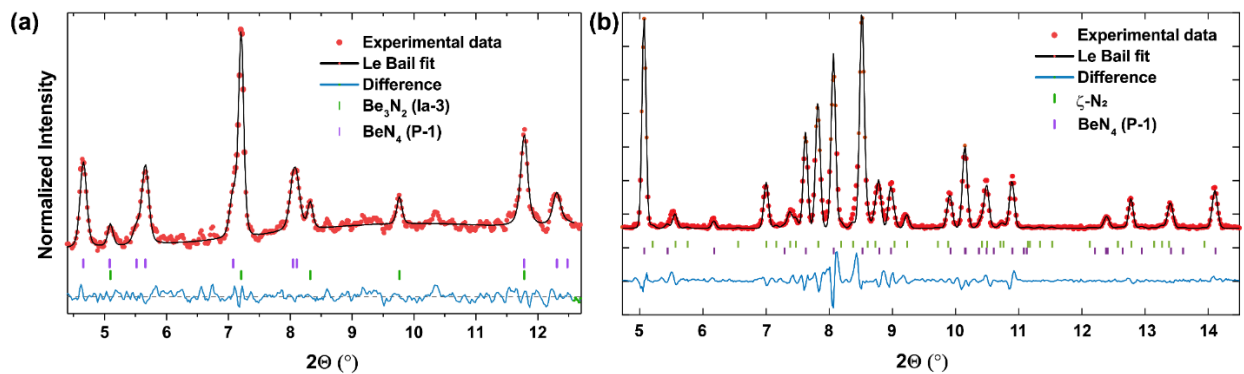

**Figure S7.** (a) Powder diffraction pattern of decompressed sample #2 containing layered *tr*-BeN$_4$ and α-Be$_3$N$_2$. (b) Powder diffraction pattern of sample #2 at the synthesis pressure (88 GPa). Wavelength 0.2952 Å for both patterns.

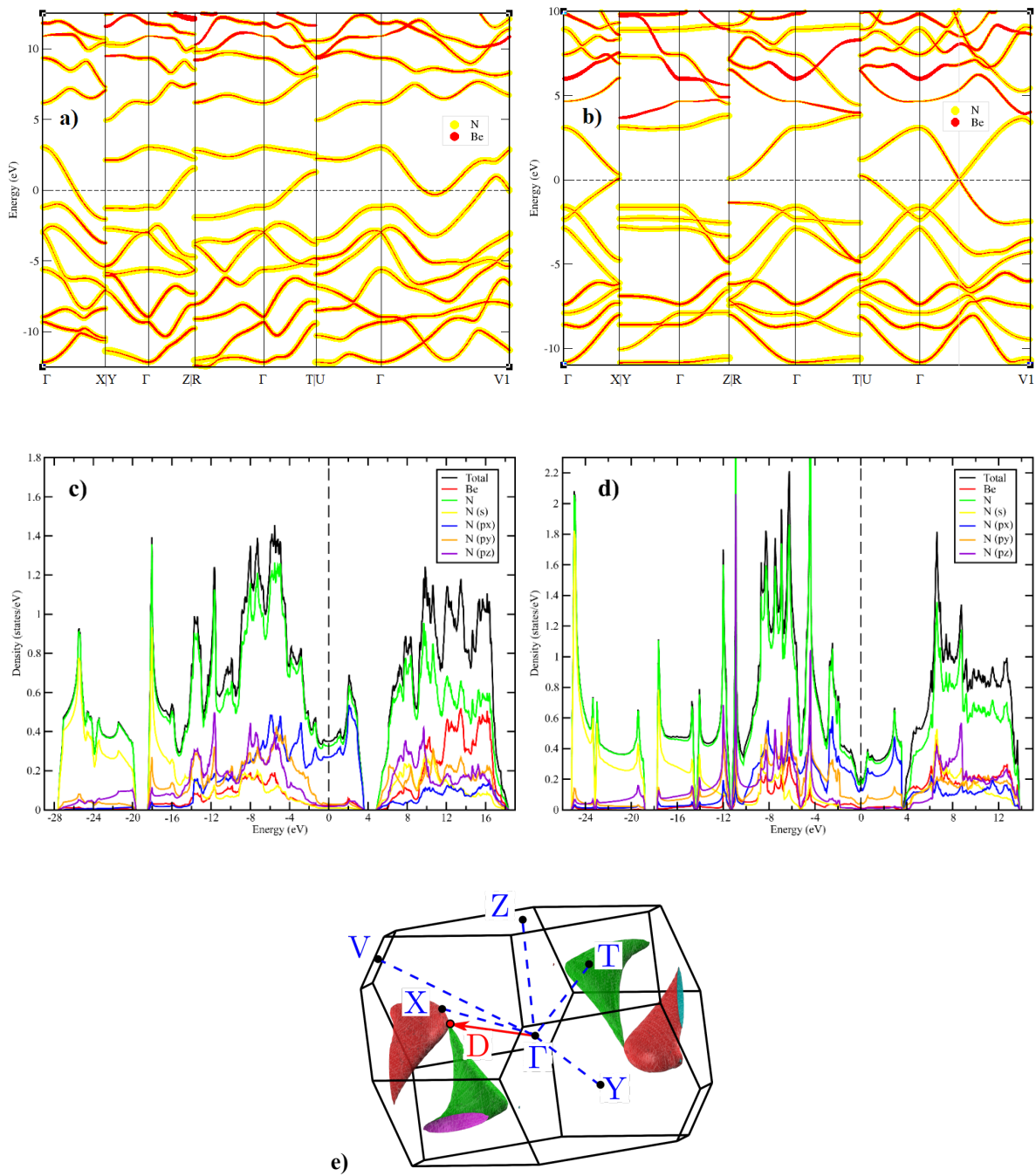

**Figure S8.** Electronic band structure (a, b) and density of states (c, d) of *tr*-BeN$_4$ at pressures P~85 GPa (a, c) and P~0 GPa (b, d), which correspond to volumes of 26.8 Å$^3$ and 43.05 Å$^3$, respectively . (e) The Fermi surface tr-BeN$_4$ at volume 43.05 Å$^3$ (pressure P~0 GPa) The position of the Dirac cone (D point) is calculated to be at (-0.37, -0.18, -0.12) in reciprocal coordinates. See Movie S2 for a close view of the anisotropy of the cone.

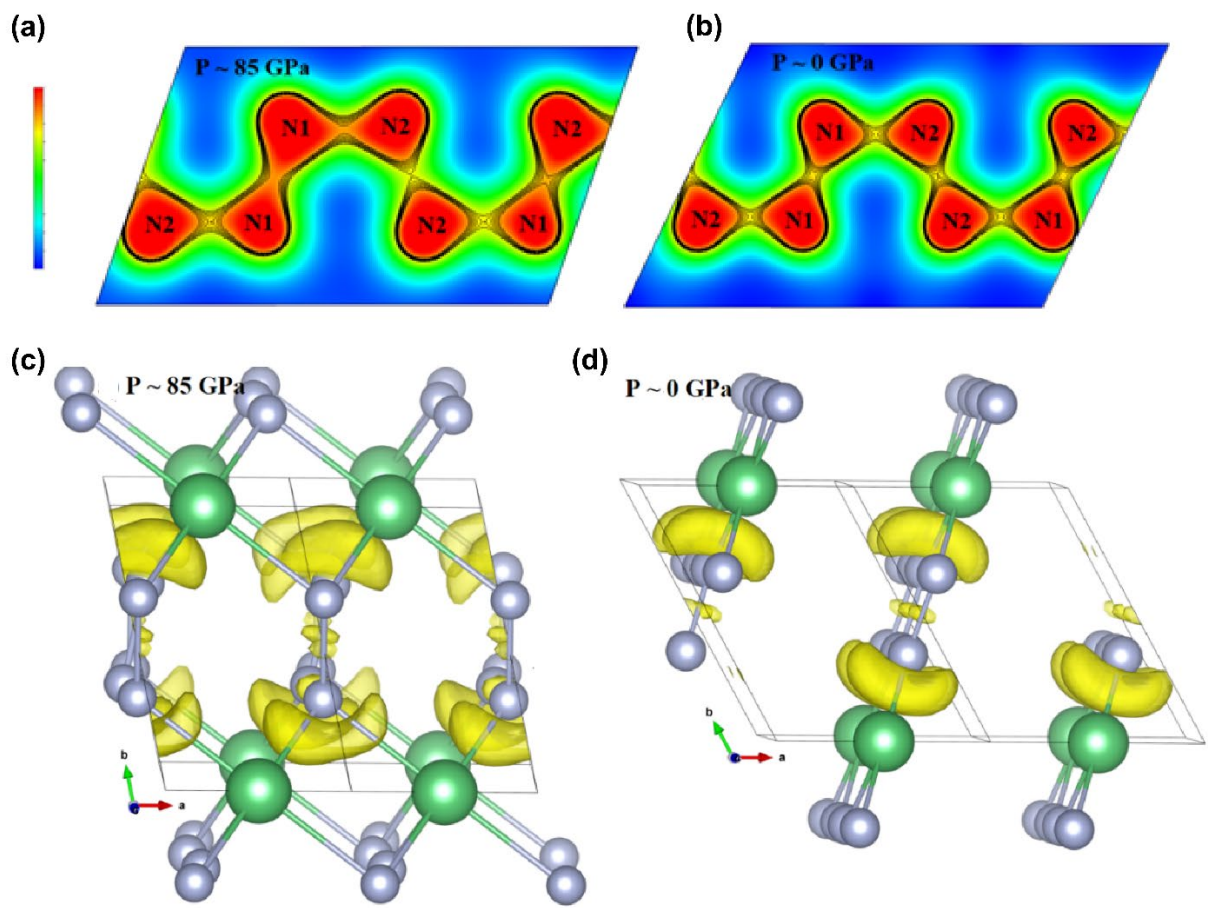

**Figure S9.** (a, b) Calculated charge density map of *tr*- BeN$_4$ at pressures P ~ 85 GPa and P ~ 0 GPa in the plane of the nitrogen chain. (c,d) Calculated electron localization function with isosurface value 0.8 (ELF) of *tr*-BeN$_4$ at pressures P~85 GPa and P~0 GPa, which correspond to volumes of 26.8 Å$^3$ and 43.05 Å$^3$. Calculated electron localization function indicates strong covalent bonding between nitrogen atoms at high and ambient pressures. Besides, at P ~ 85 GPa we can see attractors localized near Be and N2 atoms corresponding to multicenter bonds, which allow beryllium to have six-fold coordination at such a compression. The bond between the Be and N1 atoms is two-center polar covalent. Upon decompression the multicenter Be-N2-Be bonds break.

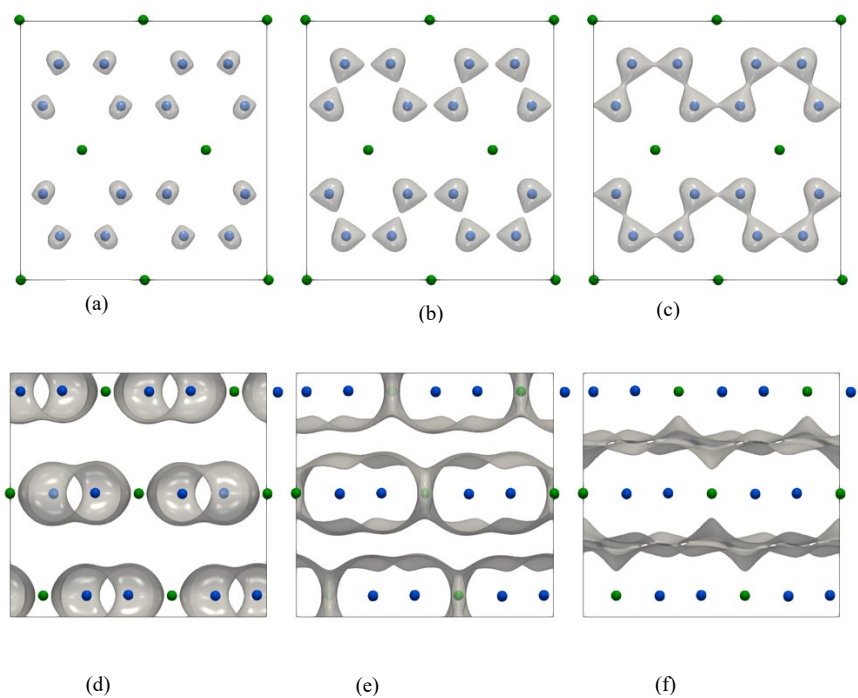

**Figure S10.** Topological data analysis of tr-BeN$_4$ charge density at zero pressure. Merge tree is constructed by following evolution of the charge density representation as the iso value decreases from high to low values. The exact value of the charge density corresponding to topological changes of the obtained iso surfaces is captured by the merge tree analysis. (a) High iso values give rise to an iso surface that appears first around the N atoms. (b) This surface grows as the iso value is lowered. (c) The first topological change occurs as the charge density iso surfaces merge into N chains. (d) As we continue to lower the iso value the iso surfaces keep expanding strictly within a BeN$_4$ layer. (e) The second topological change is observed as the charge density iso surfaces surrounding the N chains grow large enough to merge and cover the entire BeN$_4$ layer. Charge density value corresponding to this topological change will be referred to as chain separation. (f) Finally, the third topological change occurs when the iso value is so low that the iso surfaces covering individual BeN$_4$ layers merge into one. Charge density value corresponding to this topological change will be referred to as layer separation.

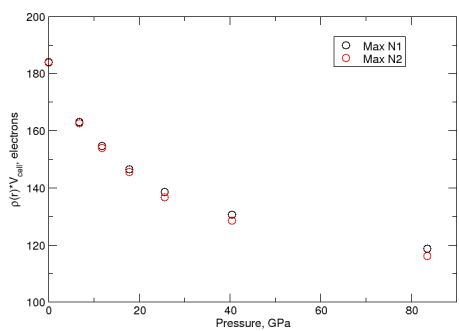 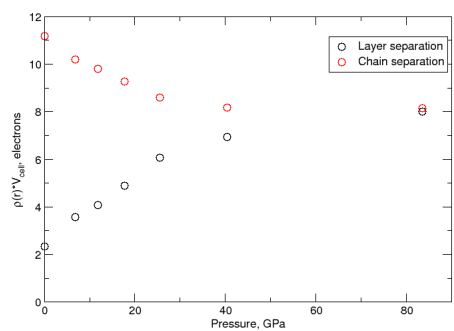

(a)          (b)

**Figure S11.** (a) The difference in the max value between two types of N atoms as a function of pressure. One can see that at synthesis pressure there is a clear difference between the two N atoms, while the atoms become similar at ambient pressure. (b) The critical charge density values corresponding to chain and layer separation (see Fig. S10 e and f, respectively) as a function of pressure. At low pressure, the values are far apart, which means that the charge density is significantly more localized within physical $BeN_4$ layers. On the contrary, at the synthesis pressure the chain and layer separation values of the charge density are close to each other, indicating a more uniform distribution of the charge density between $BeN_4$ layers.

**Table S1. Summary of syntheses in Be-N system.**

|  | Reaction mixture | Pressure, GPa* | Temperature, K | Products | Detection method |
|---|---|---|---|---|---|
| DAC # 1 | Be + $N_2$ | 41.8(5) | 2100(200) | α-$Be_3N_2$ | PXRD |
| DAC # 1 | Be + $N_2$ + $Be_3N_2$ | 60.5(5) | > 3500(300) | α-$Be_3N_2$ | PXRD |
| DAC # 2 | Be + $N_2$ | 84(4) | 2020(200) | $P$-1 $BeN_4$ | SCXRD |
| DAC # 3 | Be + $N_2$ | 98(4) | 2145(200) | $P2_1/c$ $BeN_4$ | SCXRD |
| DAC # 4 | Be + $N_2$ | 93(4) | 2000(200) | $P$-1 $BeN_4$ + $P2_1/c$ $BeN_4$ | SCXRD |

* In the experiments 2-4 pressure was determined based on the Equation of State of Re (Ref. [39]). For the pressure determination, the diffraction pattern from the very edge of the rhenium gasket was measured. According to Anzellini *et al.* [39], such pressure determination method gives pressure with ~5% accuracy.

**Table S2. Crystal structure, data collection and refinement details of *m*BeN$_4$ at 96.8 GPa**

| Crystal data | |
|---|---|
| Chemical formula | BeN$_4$ |
| $M_r$ | 65.05 |
| Crystal system, space group | Monoclinic, $P2_1/c$ |
| Temperature (K) | 293 |
| Pressure (GPa) | 96.8 |
| $a, b, c$ (Å) | 3.283(2), 3.2765(10), 4.8185(19) |
| ß (°) | 99.67 (6) |
| $V$ (Å$^3$) | 51.10 (4) |
| $Z$ | 2 |
| Radiation type | Synchrotron, λ= 0.34453 Å |
| µ (mm$^{-1}$) | 0.10 |
| Crystal size (mm) | 0.002 × 0.002 × 0.002 |
| **Data collection** | |
| Diffractometer | APS 16IDB, Pilatus 1M detector |
| Absorption correction | Multi-scan *CrysAlis PRO* 1.171.40.45a (Rigaku Oxford Diffraction, 2019) Empirical absorption correction using spherical harmonics, implemented in SCALE3 ABSPACK scaling algorithm. |
| $T_{min}$, $T_{max}$ | 0.442, 1.000 |
| No. of measured, independent and observed [$I > 2s(I)$] reflections | 102, 60, 53 |
| $R_{int}$ | 0.024 |
| $(\sin \theta/\lambda)_{max}$ (Å$^{-1}$) | 0.732 |
| **Refinement** | |
| $R[F^2 > 2s(F^2)]$, $wR(F^2)$, $S$ | 0.097, 0.229, 1.21 |
| No. of reflections | 60 |
| No. of parameters | 20 |
| Δρ$_{max}$, Δ ρ$_{min}$ (e Å$^{-3}$) | 0.71, -0.63 |

| Crystal Structure | Wyckoff Site | Coordinates ($x, y, z$) | $U_{iso}$ (Å$^2$) |
|---|---|---|---|
| Be | 2$d$ | 0.5, -0.5, 0 | 0.018(4) |
| N1 | 4$e$ | 0.095(3), -0.892(2), -0.1048(17) | 0.015(5) |
| N2 | 4$e$ | 0.238(3), -0.1947(17), -0.2559(18) | 0.010(3) |

**Table S3. Crystal structure, data collection and refinement details of trBeN$_4$ at 84 GPa**

| Crystal data | |
|---|---|
| Chemical formula | BeN$_4$ |
| $M_r$ | 65.05 |
| Crystal system, space group | Triclinic, $P\bar{1}$ |
| Temperature (K) | 293 |
| Pressure (GPa) | 83.5 |
| $a$, $b$, $c$ (Å) | 2.4062 (8), 3.5119 (9), 3.5060 (14) |
| $\alpha$, $\beta$, $\gamma$ (°) | 104.29 (3), 111.39 (3), 96.71 (3) |
| $V$ (Å$^3$) | 26.02 (2) |
| $Z$ | 1 |
| Radiation type | Synchrotron, $\lambda$ = 0.2885 Å |
| $\mu$ (mm$^{-1}$) | 0.09 |
| Crystal size (mm) | 0.002 × 0.002 × 0.002 |
| **Data collection** | |
| Diffractometer | LH@P02.2 Perkin Elmer XRD 1921 |
| Absorption correction | Multi-scan<br>*CrysAlis PRO* 1.171.40.54a (Rigaku Oxford Diffraction, 2019) Empirical absorption correction using spherical harmonics, implemented in SCALE3 ABSPACK scaling algorithm. |
| $T_{min}$, $T_{max}$ | 0.728, 1 |
| No. of measured, independent and observed [$I > 2\sigma(I)$] reflections | 167, 107, 94 |
| $R_{int}$ | 0.032 |
| $(\sin\theta/\lambda)_{max}$ (Å$^{-1}$) | 1.041 |
| **Refinement** | |
| $R[F^2 > 2\sigma(F^2)]$, $wR(F^2)$, $S$ | 0.074, 0.188, 1.17 |
| No. of reflections | 107 |
| No. of parameters | 10 |
| $\Delta\rho_{max}$, $\Delta\rho_{min}$ (e Å$^{-3}$) | 0.41, -0.48 |

**Crystal Structure**

| | Wyckoff Site | Coordinates (x, y, z) | $U_{iso}$ (Å$^2$) |
|---|---|---|---|
| Be | 1d | 0.5000 0.0000 0.0000 | 0.0085(8) |
| N1 | 2i | 0.1120(10)<br>0.6776(5)<br>0.5034(13) | 0.0066(5) |
| N2 | 2i | 0.0574(10)<br>0.6894(6)<br>0.1292(13) | 0.0069(6) |

# Supplementary References

[1] C. Prescher and V. B. Prakapenka, High Press. Res. **35**, 223 (2015).
[2] V. Petricek, M. Dusek, and L. Palatinus, Zeitschrift Für Krist. **229**, 345 (2014).
[3] G. M. Sheldrick, Acta Crystallogr. Sect. A Found. Adv. **71**, 3 (2015).
[4] G. M. Sheldrick, Acta Crystallogr. Sect. A **64**, 112 (2008).
[5] O. V. Dolomanov, L. J. Bourhis, R. J. Gildea, J. A. K. Howard, and H. Puschmann, J. Appl. Crystallogr. **42**, 339 (2009).
[6] J. Gonzalez-Platas, M. Alvaro, F. Nestola, and R. Angel, J. Appl. Crystallogr. **49**, 1377 (2016).
[7] P. E. Blöchl, Phys. Rev. B **50**, 17953 (1994).
[8] G. Kresse and J. Furthmüller, Comput. Mater. Sci. **6**, 15 (1996).
[9] G. Kresse and J. Furthmüller, Phys. Rev. B **54**, 11169 (1996).
[10] G. Kresse and D. Joubert, Phys. Rev. B **59**, 1758 (1999).
[11] J. P. Perdew, K. Burke, and M. Ernzerhof, Phys. Rev. Lett. **77**, 3865 (1996).
[12] J. Klimeš, D. R. Bowler, and A. Michaelides, Phys. Rev. B **83**, 195131 (2011).
[13] J. Klimeš, D. R. Bowler, and A. Michaelides, J. Phys. Condens. Matter **22**, 022201 (2010).
[14] A. Togo and I. Tanaka, Scr. Mater. **108**, 1 (2015).
[15] H. Carr, J. Snoeyink, and U. Axen, Comput. Geom. **24**, 75 (2003).
[16] C. Heine, H. Leitte, M. Hlawitschka, F. Iuricich, L. De Floriani, G. Scheuermann, H. Hagen, and C. Garth, Comput. Graph. Forum **35**, 643 (2016).
[17] R. F. W. Bader, *Atoms in Molecules: A Quantum Theory* (Clarendon Press, Oxford, 1994).
[18] W. Wang, S. Dai, X. Li, J. Yang, D. J. Srolovitz, and Q. Zheng, Nat. Commun. **6**, 7853 (2015).
[19] J. Wang, D. C. Sorescu, S. Jeon, A. Belianinov, S. V. Kalinin, A. P. Baddorf, and P. Maksymovych, Nat. Commun. **7**, 13263 (2016).
[20] Y. Han, K. C. Lai, A. Lii-Rosales, M. C. Tringides, J. W. Evans, and P. A. Thiel, Surf. Sci. **685**, 48 (2019).
[21] N. Marzari, A. A. Mostofi, J. R. Yates, I. Souza, and D. Vanderbilt, Rev. Mod. Phys. **84**, 1419 (2012).
[22] A. A. Mostofi, J. R. Yates, Y.-S. Lee, I. Souza, D. Vanderbilt, and N. Marzari, Comput. Phys. Commun. **178**, 685 (2008).
[23] M. I. Katsnelson, *The Physics of Graphene*, 2nd ed. (Cambridge University Press, Cambridge, 2020).
[24] N. Y. Astrakhantsev, V. V. Braguta, M. I. Katsnelson, A. A. Nikolaev, and M. V. Ulybyshev, Phys. Rev. B **97**, 035102 (2018).
[25] V. K. Dugaev and M. I. Katsnelson, Phys. Rev. B **86**, 115405 (2012).
[26] F. Aryasetiawan, M. Imada, A. Georges, G. Kotliar, S. Biermann, and A. I. Lichtenstein, Phys. Rev. B **70**, 195104 (2004).
[27] T. O. Wehling, E. Şaşıoğlu, C. Friedrich, A. I. Lichtenstein, M. I. Katsnelson, and S. Blügel, Phys. Rev. Lett. **106**, 236805 (2011).
[28] M. V Ulybyshev, P. V Buividovich, M. I. Katsnelson, and M. I. Polikarpov, Phys. Rev. Lett. **111**, 056801 (2013).
[29] J. González, F. Guinea, and M. A. H. Vozmediano, Nucl. Phys. B **424**, 595 (1994).
[30] M. Bykov, E. Bykova, G. Aprilis, K. Glazyrin, E. Koemets, I. Chuvashova, I. Kupenko, C. McCammon, M. Mezouar, V. Prakapenka, H.-P. Liermann, F. Tasnádi, A. V. Ponomareva, I. A. Abrikosov, N. Dubrovinskaia, and L. Dubrovinsky, Nat. Commun. **9**, 2756 (2018).
[31] D. Laniel, B. Winkler, E. Koemets, T. Fedotenko, M. Bykov, E. Bykova, L. Dubrovinsky, and N. Dubrovinskaia, Nat. Commun. **10**, 4515 (2019).
[32] H. E. Maynard-Casely, J. R. Hester, and H. E. A. Brand, IUCrJ **7**, 844 (2020).
[33] H. Olijnyk, J. Chem. Phys. **93**, 8968 (1990).
[34] M. I. Eremets, A. G. Gavriliuk, I. A. Trojan, D. A. Dzivenko, and R. Boehler, Nat. Mater. **3**, 558 (2004).
[35] S. Zhang, Z. Zhao, L. Liu, and G. Yang, J. Power Sources **365**, 155 (2017).
[36] J. Lin, Z. Zhu, Q. Jiang, S. Guo, J. Li, H. Zhu, and X. Wang, AIP Adv. **9**, 055116 (2019).
[37] S. R. Römer, T. Dörfler, P. Kroll, and W. Schnick, Phys. Status Solidi **246**, 1604 (2009).
[38] Y. Zhang, H. Wang, Y. Wang, L. Zhang, and Y. Ma, Phys. Rev. X **7**, 011017 (2017).
[39] S. Anzellini, A. Dewaele, F. Occelli, P. Loubeyre, and M. Mezouar, J. Appl. Phys. **115**, 043511 (2014).